\documentclass{article}
\usepackage[a4paper, total={6in, 10in}]{geometry}

\usepackage[utf8]{inputenc}
\usepackage{amsmath}
\usepackage{hyperref}
\usepackage{url}
\usepackage{graphicx}
\usepackage{booktabs}
\usepackage{lipsum}
\usepackage{array}
\usepackage[
    backend=biber,
    style=ieee,
  ]{biblatex}

\title{
Data-Driven Analysis of Text-Conditioned AI-Generated Music: A Case Study with Suno and Udio
}

\author{
Luca Casini 
\and Laura Cros Vila, 
\and David Dalmazzo,
\and Anna-Kaisa Kaila,
\and Bob L.T. Sturm
\\\\
KTH Royal Institute of Technology, Stockholm, Sweden\\
\texttt{\{casini,lcros,dalmazzo,akaila,bobs\}@kth.se}
}

\date{}

\begin{document}

\maketitle


%
%
\begin{abstract} 
Online AI platforms for creating music from text prompts (AI music), 
such as Suno and Udio, 
are now being used by hundreds of thousands of users.
Some AI music is appearing in advertising, 
and even charting, in multiple countries.
How are these platforms being used?
What subjects are inspiring their users?
This article answers these questions for
Suno and Udio using a large collection of songs generated by users of these platforms from May to October 2024.
Using a combination of state-of-the-art text embedding models, dimensionality reduction and clustering methods,
we analyze the prompts, tags and lyrics,
and automatically annotate 
and display the processed data in interactive plots.
Our results reveal prominent themes in  lyrics, 
language preference, prompting strategies, as well as peculiar attempts at steering models through the use of metatags.
To promote the musicological study of the developing cultural practice of AI-generated music we share our code and resources.\footnote{\url{https://github.com/mister-magpie/aims_prompts}}
\\\\
\textit{Submitted for revision to the Transactions of the International Society for Music Information Retrieval, ``Digital Musicology'' Special Issue} 
\end{abstract}
%


\section{Introduction}\label{sec:intro}
The application of artificial intelligence (AI) to creating media such as text, graphics and music has expanded from academic research into the commercial realm, with powerful generative tools now available to a broad public.
While obviously central for large language models like ChatGPT \cite{bubeck2023sparks}, text has become the main interface for other modalities like images, after the success of generative models like OpenAI’s DALL-E \cite{ramesh2022hierarchical} and Stable Diffusion \cite{rombach2022high}. 
Textual conditioning inputs are commonly referred to as ``prompts'' and the idea of ``prompt engineering'' is emerging around shared best practices \cite{oppenlaender2024prompting}.
AI-generated music (AI music) is starting to appear in the cultural mainstream, e.g., charting in Europe \cite{aisong_germany, aisong_norge, aisong_sweden} and appearing in marketing campaigns \cite{prezburoni}.
Professional music creators are also integrating AI into their workflows. 
A 2024 survey showed that 35\% of GEMA and SACEM members have used AI technologies in their work, reaching 51\% for those under 35 years old \cite{goldmedia2024ai}.

There are many AI-powered systems for music creation from big companies like Google, Meta, Microsoft, Open AI, Stability AI and Spotify and startups like Aiva Technologies, Boomy, Riffusion, Suno and Udio.
Given their rising popularity, how can we study the use of AI-music generation systems based on the text-to-music interaction modality?
What large-scale patterns can we observe? What can we learn about the users of these systems and their goals?
This paper focuses on analyzing  AI-generated music from the platforms Suno and Udio, as they represent the most popular services to offer text-to-music generation, with hundreds of thousands of users \cite{Yang2025} and active communities on social media like Reddit and Discord.\footnote{At the time of writing Suno's discord has $397,642$ members and Udio has $17,435$.}

We approach these musicological questions with a data-driven methodology, using tools from natural language processing and data science.
This methodology allows one to manage the unique nature of this new music practice by shifting from in-depth analysis of individual works to a broader, data-driven exploration of music produced by a broad community, thereby uncovering large-scale trends in AI music.
We start by collecting a large dataset of over $100,000$ songs generated 
by users of Suno and Udio from May to October 2024.
This dataset consists of textual prompts, lyrics and tags
associated with the music generated by these platforms.
In the course of our exploration, another dataset of AI music appeared:
the SONICS dataset \cite{rahman2024sonics}---but this dataset is collected for studying the problem of AI music detection,
and is not representative of how Suno and Udio are being used.
We perform exploratory data analysis of the textual metadata in our dataset
to discover patterns and trends.
This allows us to answer questions tied to demographic aspects such as what languages and alphabets are the most prevalent.
Through textual analysis and clustering we can answer questions about prompts and lyrics: What emerging prompting strategies can we observe?
What type of qualifiers appear most often in the style prompt?
What are the most common themes that appear in the lyrics?
How many prompts reference real musicians? Are users trying to impersonate real artists?

Despite the growing prominence of AI music generation platforms, there has been no systematic analysis of how these systems are actually being used in practice \cite{sturm2024ai}. 
This work aims to address this lack and promote the emerging field of AI music studies.
This paper presents three main contributions.
We develop and apply a new methodology for the study of AI music from text-controlled systems based on textual features. We do this by combining well established techniques and state-of-the-art tools from related fields and putting them to the test for this specific domain.
To facilitate reproducibility, future research and comparative studies, we share our code along with a list of URLs for the songs in our dataset.\footnote{\url{https://github.com/mister-magpie/aims_prompts}}
Finally, we produce interactive visualizations that can help explore and navigate the dataset we assembled.\footnote{An interactive version is available online at \url{https://mister-magpie.github.io/aims_prompts/}}

The article is structured as follows:
Section \ref{sec:related} provides background information on lyrics and prompt analysis,
and Section \ref{sec:interaction} describes the interaction modality of these platforms, 
and explains how they integrate metadata in the generation process.
Section \ref{sec:methodology} details the methodology of our study.
Section \ref{sec:dataset} describes the dataset we collect and use for our analyses.
Section \ref{sec:results} illustrates the results we obtain from our analyses of prompts, lyrics, tags and metatags. 
Section \ref{sec:discussion} contains a discussion of the results and points to further work.
Finally, Section \ref{sec:conclusion} concludes the paper.

\section{Related Work}\label{sec:related}
We first review work done in prompt analysis, particularly in the context of text-to-image generation. 
Then we present research on lyrics analysis, focusing on computational methods for extracting meaning, sentiment, and stylistic patterns from song lyrics.

\subsection{Prompt Analysis}
To the best of our knowledge there is no work that studies prompting in the context of music generation.
For this reason we turn to similar works in the context of image generation.
A notable example is the work by \cite{sanchez2023examining}, investigating text-to-image generation practices by looking at \textit{DiffusionDB} \cite{wangDiffusionDBLargescalePrompt2022}, a dataset of 14 million real images and prompts from real users of Stable Diffusion. 
Their analyses result in a taxonomy of specifiers found in the prompts and clusters to find macro categories, along with a classifier trained to follow it.
An extension of this work comes from \cite{artstation}, who include data from \textit{Midjourney} and provide some additional clustering and topic modeling.
Both papers make use of the HDBSCAN clustering algorithm in conjunction with UMAP dimensionality reduction. 


\subsection{Lyrics Analysis}
In MIR tasks such as mood and genre classification, lyrics and other textual features have been analyzed using natural language processing (NLP) techniques \cite{Mayer2011, Mayer2010b}.
%
In the context of genre classification, \cite{logan2004semantic} demonstrate how to measure song similarity using NLP techniques applied to lyrics.
%
\cite{fell2014lyrics} use a number of specialized linguistic features to address genre classification, rating prediction and publication time prediction.
%
\cite{pyrovolakis2022multi} use a mix of NLP models and audio signal processing techniques to create song mood classifiers.

Similar techniques can also be used to investigate sociocultural questions, moving beyond classification and retrieval tasks.
For example, \cite{varnum2021song} looked at the decreasing linguistic complexity of lyrics in six decades of popular American music, using compressibility as a proxy, and studied correlation to ecological, social and cultural factors in order to explain it.
\cite{napier2018quantitative} apply sentiment analysis techniques to lyrics of popular music since the 1950s and show how certain feelings like anger, disgust, fear, sadness, and conscientiousness have increased significantly, while joy, confidence, and openness expressed in pop song lyrics have declined.
In their analysis of lyrical complexity, structure, emotion, and popularity across five decades, \cite{parada2024song} similarly demonstrate an increase in negative emotions, as well as a decrease in both lexical and structural complexity over time.  

\section{The interaction modality of Text-to-Music generation systems}\label{sec:interaction}
\begin{figure}[t]
    \centering
    \includegraphics[height=12cm]{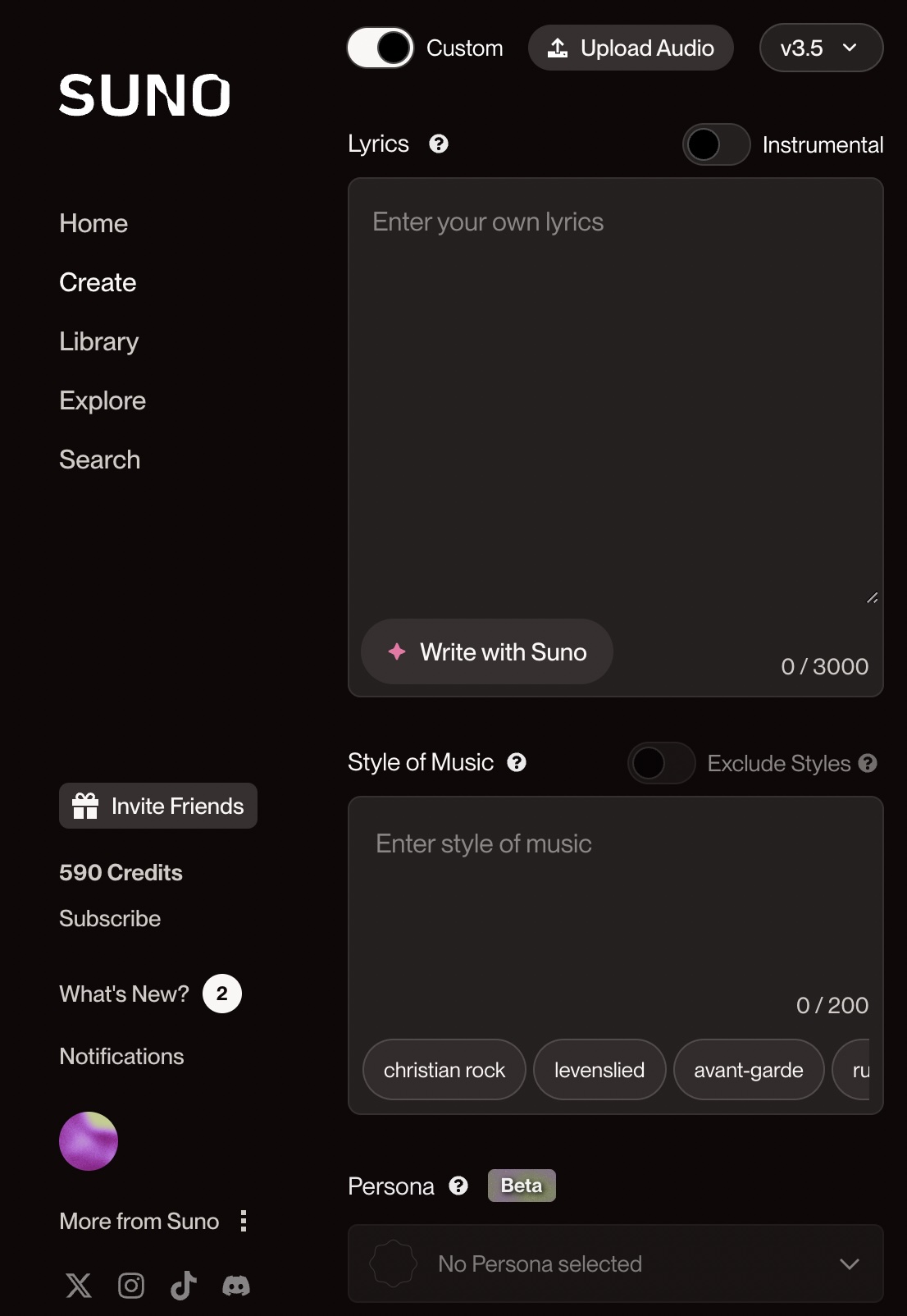}
    \caption{{\em Suno} website interface when ``Custom'' mode is selected. 
    Here the user can input the lyrics as well as the style of the music. Last accessed Dec. 16, 2024.}
    \label{fig:suno_interface}
\end{figure}

Figure \ref{fig:suno_interface} shows the music creation interface in Suno.
To start generating a song, the user can click on the ``Create'' button in the sidebar.
This will show a textbox for prompting the AI model.
Lyrics and style tags will be generated according to the prompt and used as conditioning for the model. 
Users can also specify, using a radio button, if they prefer an instrumental track.
An additional radio button labeled ``Custom'' can be toggled, offering direct control over lyrics and style prompts. 
While the interface suggests tags that are added as comma-separated elements, the textbox also accepts free text.

\begin{figure}[h]
    \centering
    \includegraphics[height=12cm]{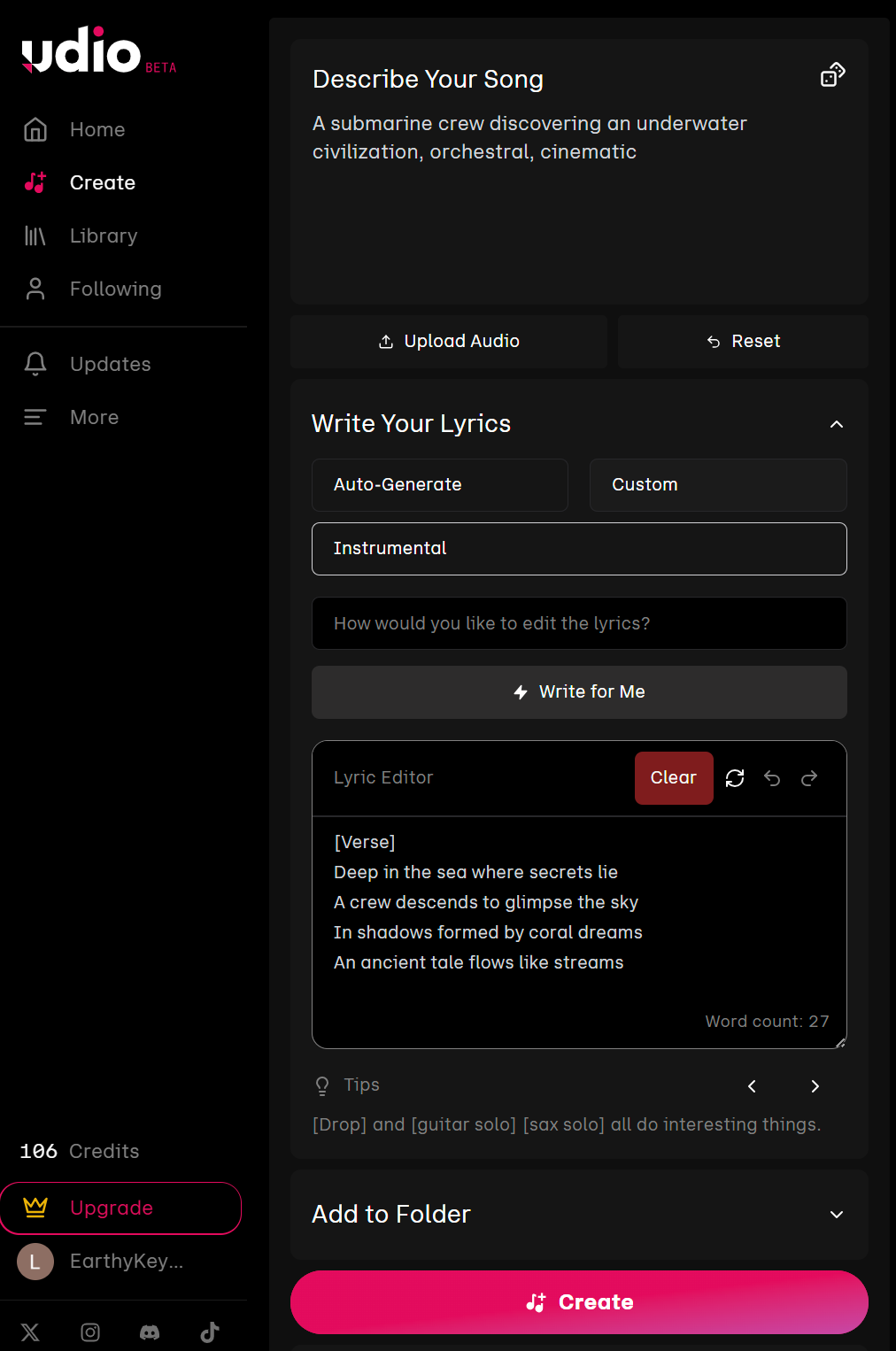}
    \caption{Creation interface for Udio (as of Feb 18th 2025). Selecting the Custom allows for manual adjustments even if the prompts are automatically generated.}
    \label{fig:udio_interface}
\end{figure}

Figure \ref{fig:udio_interface} shows an example of the generation interface in Udio.
The interaction happens in largely the same way as Suno.
The user is first asked to describe the song they want and popular tags are suggested.
There is a button to fill this part with a randomly generated prompt.
After that, lyrics can be written manually or automatically or skipped to generate an instrumental song.
The interface suggests pressing the \texttt{/} key to show a list of popular meta-tags that are added to the lyrics between square brackets and allow for additional control.

The text fields in the interfaces accept any Unicode input, 
including emojis and non-Latin characters.
At the time of writing, new functionalities are being added to the interfaces of both platforms (e.g., audio prompting, new models);
but we do not consider any additions beyond the last  
date we collected (October 2024).

\section{Methodology}\label{sec:methodology}

We now describe our methodology,
drawing on similar work analyzing AI generated images \cite{artstation,sanchez2023examining}. 
Our method can be summarized in six steps.
\begin{enumerate}
    \item We scrape the regularly updated playlists from Suno and Udio's ``new songs'' public-facing homepage. 
    \item We extract the metadata from these playlists (prompt, lyrics, and tags) and cast it as a dataframe consisting of rows (songs) and columns (metadata). This allows us to filter the rows based on content (e.g., language, length, null values).
\end{enumerate}
We perform separate analyses on prompts, tags and lyrics. 
For any of those datatypes:
\begin{enumerate}
    \setcounter{enumi}{2}
    \item We embed it using a pretrained model such as \textit{NV-Embedv2} \cite{lee2024nv}.
    \item We perform dimensionality reduction and clustering on the embeddings.
    \item For each of the clusters, we derive a label using an automated procedure and manually refine. 
    \item We create a 2D scatter plot of the clustered elements for visual analysis.
\end{enumerate}
We iterate steps three to six multiple times, 
experimenting with the hyperparameters for dimensionality reduction and clustering.

In more detail, we perform scraping using a script that extracts metadata from the HTTP request we find in the homepage code for both platforms.
We save the metadata in JSON format for each song in a file which is  refined into a dataframe for each platform.
Some of the analyses we perform require filtering the dataset.
We remove some entries only in those specific cases.
Specifically, this means ignoring languages other than English from both the lyrics and prompt subsets, removing entries that are too short or too long, and dropping empty rows from the dataframe.
We use a language classification model to filter out non-English elements 
because these could prove problematic for English-tuned NLP tools.
We choose the classification model included in the \texttt{fastext} library from Meta.\footnote{\url{https://huggingface.co/facebook/fasttext-language-identification}}
The model outputs labels for languages according to the ISO 639-3 standard\footnote{\url{https://en.wikipedia.org/wiki/ISO_639:y}} together with an indication of the alphabet used.

To generate embeddings for clustering, we use two state-of-the-art deep learning models specifically designed for text embedding. We use \textit{NV-Embedv2} by NVIDIA \cite{lee2024nv}.
This model uses decoder-only transformers along with a modified attention mechanism and ad-hoc contrastive learning pre-training to learn text embeddings, outperforming previous methods based on masked language models.
We employ the implementation made available by the authors through the Huggingface library \footnote{\url{https://huggingface.co/nvidia/NV-Embed-v2}}.

The latent representation produced by NV-Embed has a dimensionality of $4096$. 
We use dimensionality reduction as an intermediate step before clustering.
Our choice for this task is \textit{UMAP} \cite{mcinnes2018umap},
which works by building a high-dimensional graph with weighted connections that is reduced to a smaller dimensional graph while trying to preserve its structure.
We employ the python implementation provided by the \texttt{umap-learn} package\footnote{\url{https://umap-learn.readthedocs.io/}}.
UMAP has a number of important parameters that influence the final result.
The number of approximate nearest neighbors when building the high-dimensional graph regulates the trade-off between global and local structure, and this in turn influences the outcomes of clustering in our pipeline.
The minimum distance between points in the reduced dimensional space directly affects how closely the algorithm packs together similar points.
At clustering time, this can lead to too many clusters if set too low, while at visualization time it can result in small clusters being too far away from the rest, making the plot difficult to read.

For clustering, we use
\textit{HDBSCAN} \cite{campello2013density}, which is an extension of the DBSCAN algorithm that enables robust identification of clusters with varying densities \cite{ester1996density}.
This clustering algorithm is recommended by \cite{mcinnes2018umap}, in conjunction with UMAP, when operating in high-dimensional embedding spaces.
One benefit of \textit{HDBSCAN} compared to other clustering algorithms is the automatic detection of outliers.
We use the implementation in the \texttt{scikit-learn} package\footnote{\url{https://scikit-learn.org}}.
The algorithm has hyper-parameters for the minimum and maximum cluster size, as well as a factor $\epsilon$ that controls how likely clusters are to be merged together.
We set these values for each analysis through trial and error, aiming for the best combination of hyper-parameters that gives a meaningful number of clusters.

To facilitate exploration, especially in the case of lyrics with a high number of clusters, we devise a strategy to automatically name them.
After forming clusters of tags, we use a bag-of-words strategy by counting term prevalence and then select the $N$ most common ones.
We find that $N=3$ is enough to topically characterize clusters.
For lyrics and prompts, we remove stop-words, perform lemmatization and finally select the top-3 words after TF-IDF ranking using the \texttt{spacy} package. 
We manually inspect clusters and then possibly rename each with more descriptive labels.
We find, however, that the top word often provides a good characterization of the cluster.
Additionally, we assign tags to a set of macro categories by either using a list of genres or instruments, or by running regular expressions on specific technical terms (e.g., \texttt{BPM}).
This influences dimensionality reduction and clustering, making the final clusters more informative and reflective of higher-level categories. 

\section{Dataset}\label{sec:dataset}
Our dataset consists of a total of $101,953$ songs --- $20,519$ from Udio and $81,434$ from Suno --- 
collected between May and October 2024.
%
Table \ref{tab:metadata_fields} summarizes the columns relevant for our study out of all those present in the metadata JSON object we have.
\begin{table}[htb]
\centering
\footnotesize
\begin{tabular}{m{2.2cm}m{2.4cm}m{2.4cm}}
\toprule
\textbf{Field}& \textbf{Suno}& \textbf{Udio}\\ 
\midrule
\texttt{id} & Unique song ID & Unique song ID \\ \hline
\texttt{title} & Title of the song & Title of the song \\ \hline
\texttt{tags} & Tags derived from the user input & User tags \\ \hline
\texttt{replaced\_tags} & N/A & dictionary with tags replacements \\ \hline
\texttt{lyrics} & N/A & Song lyrics \\ \hline
\texttt{prompt} & Song lyrics & User input \\ \hline
\texttt{gpt\_description \_prompt} & User prompt for lyrics generation & N/A \\ \hline
\texttt{optimized\_prompt} & N/A & Refined user input \\
\bottomrule
\end{tabular}
\caption{The subset of the metadata  from Suno and Udio we analyze}
\label{tab:metadata_fields}
\end{table}

There are inconsistencies around the actual meaning of \texttt{prompt} and \texttt{tags} due to differences between the two platforms. 
In the context of Suno, \texttt{prompt} means any text input in the lyrics textbox before having it generate a song (Fig. \ref{fig:suno_interface}). 
This text can be written either by a user, or generated by an LLM prompted with a user's description, stored as \texttt{gpt\_description\_prompt}. 
In Suno, what gets labeled as \texttt{tags} in the metadata comes from a textbox labeled in the interface as \textit{Style of Music}, with a popover that invites the user to:
\textit{``Describe the style of music you want (e.g., acoustic pop). Suno's models do not recognize artists' names but do understand genres and vibes.''}. 
Although suggestions are provided, resulting in comma-separated keywords if clicked, the string does not appear to be post-processed and often results in very long individual tags. The way that these are stored is reflected in the interface too.
Udio offers a textbox for a general prompt, which is then translated into tags by the system, unless \textit{Manual Mode} is enabled. 
Additionally, a number of suggested tags is provided, which updates as the user selects them or writes in the textbox.
The post-processing of tags is most notable when the name of a real artist is mentioned; these get replaced with generic tags, as opposed to Suno, which just blocks their use.
The metadata contains the original prompt and tags, as well as their post-processed version, if any.
The interface has a section for lyrics, which can be either AI- or user-generated.
In both cases they are stored as \textit{lyrics} without any indication of their provenance.
The \textit{optimized prompt} field for Udio is used inconsistently. 
When present, it seems to overlap with the \textit{replaced tags} field.

In the context of this paper, we use the following terminology in order to clear the ambiguity around certain fields of our metadata: 
\begin{description}
    \item[lyrics:] \texttt{prompt} in the case of Suno and \texttt{lyrics} in the case of Udio; 
    \item[tags:] \texttt{tags} and their replaced/optimized version in the case of Udio and high level labels for each song
    \item[prompt:] \texttt{gpt\_description\_prompt} in the case of Suno and \texttt{prompt} in the case of Udio
\end{description}

\section{Results}\label{sec:results}
We now detail the specifics of several analyses and discuss the results. 

\subsection{Language Detection}\label{sec:lang_detec}
Table \ref{tab:lang_dist} shows the $15$ most prevalent languages in lyrics on both platforms, accounting for more than 90\% of the dataset.
For each language, prevalence in each platform is given in percentage points, with its rank depending on their sum across platforms
\begin{table}[htb]
\centering
\begin{tabular}{m{2cm}m{2cm}rr}
\toprule
\textbf{Language} & \textbf{ISO 639-3} & \textbf{Udio} & \textbf{Suno} \\
\midrule
English     & \texttt{eng\_Latn} & 71.39\% & 46.75\% \\
German      & \texttt{deu\_Latn} &  3.68\% &  8.87\% \\
Russian     & \texttt{rus\_Cyrl} &  2.99\% &  6.68\% \\
Spanish     & \texttt{spa\_Latn} &  3.28\% &  4.58\% \\
Portuguese  & \texttt{por\_Latn} &  1.68\% &  3.55\% \\
Korean      & \texttt{kor\_Hang} &  3.21\% &  3.00\% \\
Chinese     & \texttt{yue\_Hant} &  1.77\% &  3.33\% \\
Indian      & \texttt{ind\_Latn} &  0.27\% &  3.26\% \\
French      & \texttt{fra\_Latn} &  1.81\% &  2.15\% \\
Japanese    & \texttt{jpn\_Jpan} &  1.45\% &  1.92\% \\
Turkish     & \texttt{tur\_Latn} &  0.29\% &  1.66\% \\
Italian     & \texttt{ita\_Latn} &  1.06\% &  1.29\% \\
Thai        & \texttt{tha\_Thai} &  0.05\% &  1.26\% \\
Vietnamese  & \texttt{vie\_Latn} &  0.09\% &  1.11\% \\
Polish      & \texttt{pol\_Latn} &  0.77\% &  0.94\% \\ \midrule
TOTAL	    &                    & 93.79\% & 90.35\% \\
\bottomrule
\end{tabular}
\caption{The 15 most popular languages for lyrics in the dataset. The percentage is referred to the prevalence in each platform.}
\label{tab:lang_dist}
\end{table}
For both services, we see prompts written in a variety of languages, with English being by far the most prevalent.
This is to be expected as both companies are based in the USA.
The tail of the distribution features the major European and Asian languages.
The differences in use of certain languages might be explained by different release patterns on the global market. 
Additionally, the user-base of Udio appears to be smaller than Suno, which may be due to its more recent establishment.

\subsection{Prompts}\label{sec:promtps}

In our dataset of $101,953$ songs, we find $41,108$ that have a prompt: $20,589$ from Suno and $20,519$ from Udio.
The median value is around $80$ characters, but for Udio, there is a heavier tail of long prompts pushing the $99^{th}$ percentile to $1,325$ characters.
Using this number as a threshold, we remove $206$ prompts based on their length.
To avoid distorting the results because of different languages and repeated prompts, we filter out any text in languages different than English and remove duplicates.
This results in a subset of $16,881$ prompts.

\begin{figure*}
    \centering
    \includegraphics[width=\linewidth]{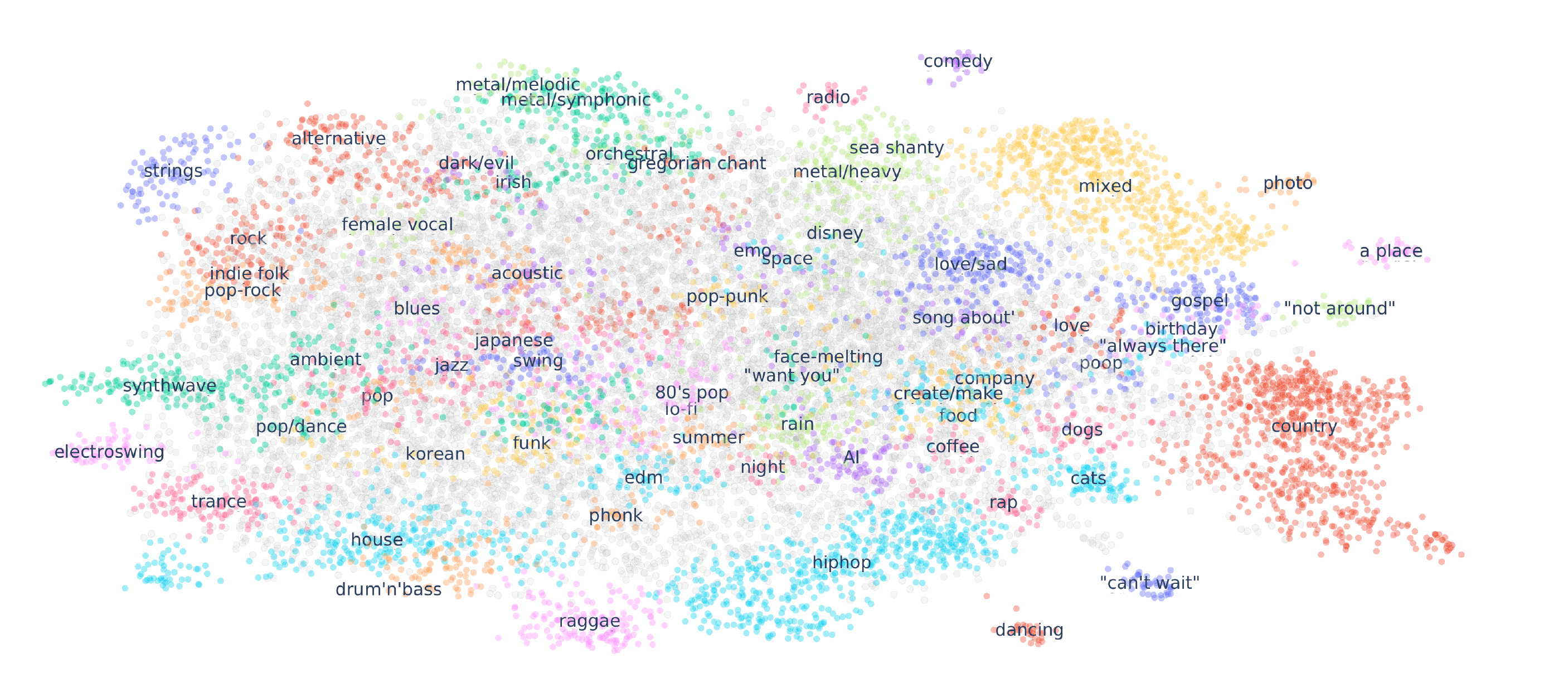}
    \caption{Clustering of prompts embeddings. The names for each cluster are manually defined after checking their content.}
    \label{fig:prompts_nvembed}
\end{figure*}

We use \textit{NV-Emebed v2} to embed the prompts and reduce the resulting embedding to five dimensions using UMAP with \texttt{n\_neighbors}$=10$ and \texttt{min\_dist}$=0.15$.
We run \textit{HBDSCAN} with a minimum cluster size of $20$ and maximum of $200$, with $\epsilon=0.25$, resulting in $81$ clusters.
Figure \ref{fig:prompts_nvembed} shows the resulting plot.
For each cluster in the plot, its label is manually reviewed and revised based on the 
procedure described in Sec. \ref{sec:methodology}.

We find that the clusters form around shared keywords and semantic meaning.
Most of the clusters seem to build around genre- or instrument-specific prompts.
We also see a number of clusters built on themes like ``pets'', ``coffee'', ``summer'' or ``stand-up comedy''.
A few clusters have names between quotes; these are clusters that feature prompts where some details might change but the part between quotes is always present.
For example, ``An xxx song about when your not around'' (sic).

An interesting group is made of prompts that appear to be generated automatically, possibly by some external service.\footnote{e.g., \url{https://sunoprompt.com}, last accessed Mar. 12 2025.}
We did not include these in Fig. \ref{fig:prompts_nvembed} as their high degree of similarity results in a skewed plot.
These are labeled ``scripted'' in the interactive visualization.\footnote{\url{https://mister-magpie.github.io/aims_prompts/}}
As the name suggests, these prompts follow a pattern that features a theme, a name the song is dedicated to, some adjectives, and a genre.
As a confirmation of their scripted nature, the prompts feature a string between curly brackets that was likely not replaced by a value, e.g., ``I'm currently feeling hyped and I want to feel happy. My favorite genre of music is country. Please write the song in \texttt{{moodsGenre}} style.''

If we cluster prompts without removing duplicate tags, we also obtain a number of clusters containing only duplicates of the same prompt, which shows that some users publish multiple attempts at generating the same song.
We can speculate that most of the generations on the platform are of this nature but since they remain unpublished we cannot be confirm this. 

Beyond semantic content, it appears that there are two main tendencies in the way prompts are built.
On the one hand, we find prompts constructed like a list of comma-separated qualifiers, with the length varying between a couple of words to over ten.
(e.g., ``modern country, contemporary folk, introspective, melodic, bittersweet'').
On the other, we have more literary prompts that describe the content of the song 
(e.g., ``A song about \dots'') 
or the sound of the song 
(e.g., ``A jazz ballad with trumpet, etc., \dots'').
We can view these two as extremes on a spectrum, with most prompts being a combination of both approaches.

Prompts might sometimes contain references to real artists, with text like ``A song in the style of \dots'' or ``Sung by \dots''.
It appears, however, that Suno uses the prompt only for generating the lyrics, while tags do not allow the use of real artists names. 
Udio is instead more flexible, extracting tags from the prompt automatically and in many cases allowing artists' names.
Section \ref{sec:artists_names} contains a list of artists we find in the Udio subset.
Overall, it appears that prompts contain a mix of information about themes and style.
The former emerges more clearly from analyzing lyrics, while the latter is better observed in the tags.

\subsection{Lyrics}\label{sec:lyrics}
In our dataset, we find almost $11\%$ (Udio) and $15\%$ (Suno) are
instrumental, and thus have no or few lyrics.
We filter all lyrics that are not detected as English.
Looking at the distribution of lyrics character length shows 
that the $99^{th}$ percentile is $3,460$ for Udio and $2,999$ for Suno. 
We thus remove from consideration English-detected lyrics longer than $3,600$ characters
and instrumental entries, giving a total size of $42,163$ lyrics that we analyze.

We combine song lyrics and song title into a single string for analysis.
Since lyrics can include structural information inside square brackets, e.g., ``[chorus]'', 
we remove all such occurrences using a regular expression.
Following \cite{artstation}, we reduce the embeddings of the lyrics to five dimensions using \texttt{umap-learn} default parameters (except the \texttt{metric} setting that was set to \texttt{cosine}).
%
We then apply the HDBSCAN in the reduced latent space.
After some experimentation, we decide to use default settings and set the minimum cluster size to $20$.
The algorithm gives us $190$ clusters, with cardinalities ranging from $20$ to $854$, and a group of outliers of size $25,794$.
This high number of outliers is not problematic since our goal is not to classify all the elements perfectly but to find meaningful anchors 
in the latent space with which to infer what each cluster represents.

\begin{figure*}[htb]
    \centering
    \includegraphics[width=\textwidth]{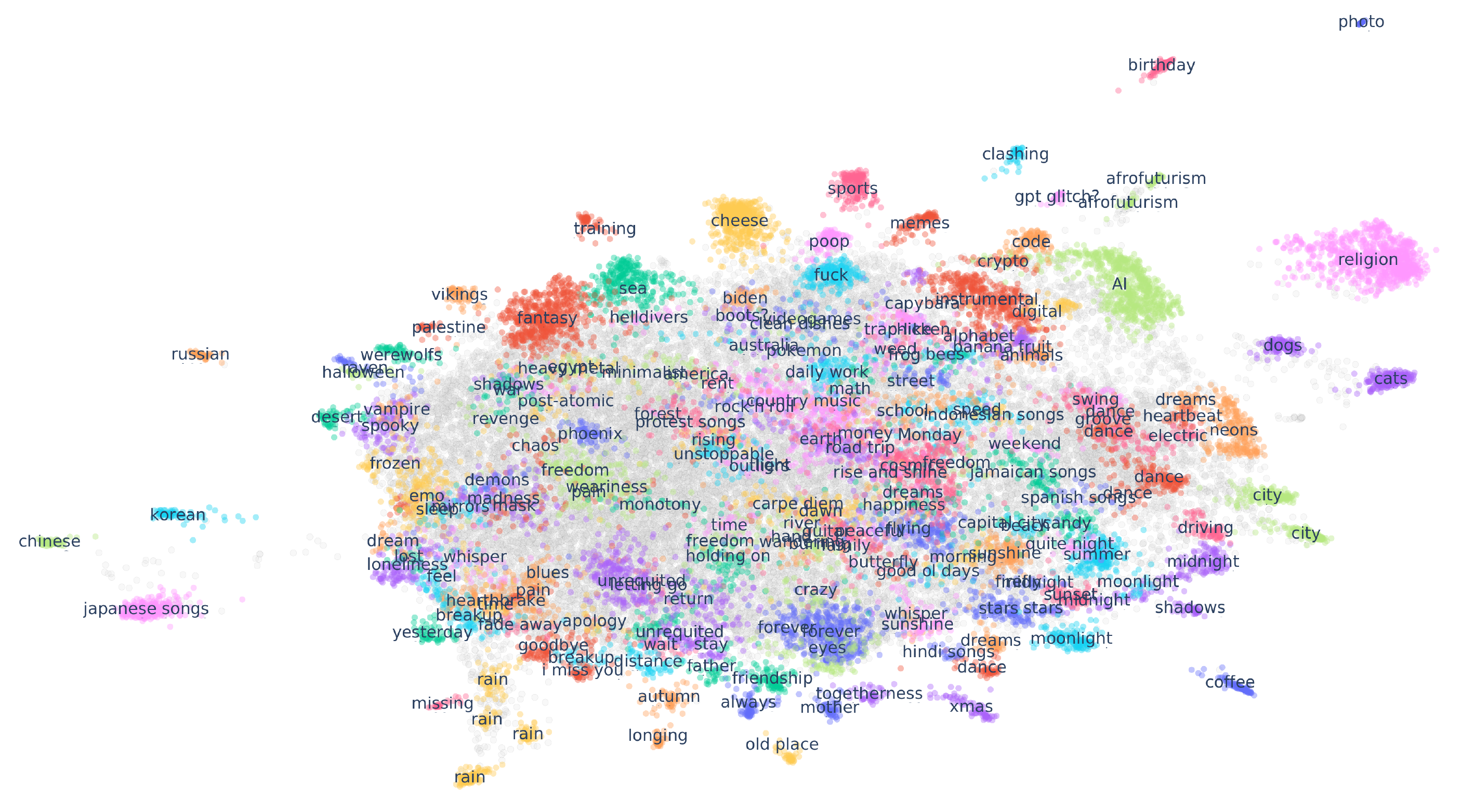}
    \caption{Clusters of lyrics (from both Suno and Udio) obtained from the HDBSCAN algorithm applied to a 5-dimensional reduction of the embedding space obtained from UMAP. The cluster is then visualized on a 2-dimensional reduction of the same space. Colors represent clusters, whose name is printed at the centroid.
    }
    \label{fig:lyrics_umap}
\end{figure*}

We manually refine the auto-generated cluster names by listening to the songs closer to the centroid of each cluster and renaming the cluster in cases where we find the original name imprecise or ambiguous.
We then manually group clusters into macro-categories 
based on shared semantic similarities.
We obtain $26$ macro-categories, plus one for the outliers.
Table \ref{tab:lyrics_cat} summarizes this result
and Fig- \ref{fig:lyrics_umap} shows the bi-dimensional reduction of the embedding space, where each song is a dot colored according to cluster whose name is printed near the centroid.

\begin{table}[htb]
\centering
\footnotesize
\begin{tabular}{{m{1.5cm}m{6cm}}}
\toprule
Category & Cluster \\ \midrule
abstract & clashing, afrofuturism, dreams, religion, sleep, chaos, flying, post-atomic, shadows, carpe diem, money, war, mirrors, mask, photo \\ \hline
animals & frog, butterfly, bees, firefly, capybara, chicken, animals, dogs, cats \\ \hline
celebration & birthday, xmas, halloween \\ \hline
daily life & weekend, monotony, school, rent, Monday, daily work, clean dishes \\ \hline
dance & dance, swing, moonlight, heartbeat, groove \\ \hline
driving & speed, driving, road trip \\ \hline
family & friendship, mother, father, family \\ \hline
fantasy & werewolfs, vikings, demons, shadows, fantasy, vampire, spooky \\ \hline
feelings & old place, revenge, freedom, good ol days, pain, peaceful, happiness, yesterday, weariness, madness, loneliness, fade away \\ \hline
food & coffee, candy, cheese, banana fruit \\ \hline
genre & guitar, rock'n'roll, blues, country music, trap-like, heavy metal, emo \\ \hline
location & desert, capital city, egypt, earth, beach, river, america, australia, forest, sea \\ \hline
love & wandering, longing, holding on, distance, time, burning, stay, light, missing, i miss you, breakup, crazy, return, feel, togetherness, always, unrequited, electric, whisper, hand, letting go, apology, goodbye, wait, hearthbrake, dream, forever, eyes \\ \hline
meme & weed, memes, poop, fuck \\ \hline
languages & russian, korean, chinese, japanese, indonesian, hindi, jamaican, spanish \\ \hline
motivational & rise and shine, dreams, dawn, phoenix, unstoppable, rising \\ \hline
other & raven, gpt glitch, alphabet, instrumental, boots?, minimalist?\\ \hline
outliers & outliers \\ \hline
politics & palestine, biden, protest songs \\ \hline
weather/ seasons & autumn, summer, rain, frozen, moonlight, sunshine \\ \hline
sports & sports, training \\ \hline
stars/night & quite night, stars, cosmic \\ \hline
technology & math, crypto, digital, AI, code \\ \hline
time & midnight, sunset, time, morning \\ \hline
urban & city, neons, lost, street \\ \hline
videogames & pokemon, helldivers, videogames \\
\bottomrule
\end{tabular}
\caption{Macro categories and clusters in the lyrics embedding space}
\label{tab:lyrics_cat}
\end{table}

We see certain types of use and users emerge from the clusters.
As one might expect, the biggest group is songs with variations around the theme of love with many smaller clusters that capture the different aspects like: unrequited love, breakups, distance, longing, etc.
The biggest individual cluster, however, is the one with $854$ number of worship songs, where \texttt{love, god} and \texttt{lord} are the most prevalent words. 
In this cluster, Christianity and Islam seem to be the dominant subjects.
There is a group of songs that are created to celebrate a specific event like holidays, or family members. 
This suggests that users sometimes use these services to generate bespoke songs for particular people on celebratory occasions, or in relation to certain experiences like trips or sport events, or daily routines like school life.
Songs about pets, animals and food are also a recognizable category.
These clusters demonstrate an unprecedented ``musicalization of everyday life'' as noted by \cite{tan2024we}.
%
Current geopolitical events are also reflected on the platform.
We find a cluster of political songs about the 2024 USA elections and candidates.
Clusters for the Israeli-Palestinian conflict are also present, but in the plot we only see a cluster with Palestine-related keywords. Songs with Israel-related keywords are not absent from dataset but rather they are mostly labeled as Hebrew by the language detection and thus are not considered in our analyses.
%
Not all traces of languages other than English are filtered out, however. 
A subset of users seems to be creating songs using bilingual lyrics. 
This is especially true for Asian languages, likely because of the popularity of J-Pop and K-Pop.
%
Themes from fiction like high fantasy, horror and monsters, and post-atomic worlds all appear with their clusters of songs, with the first being the largest.
There are multiple small clusters about video-games, with classics like \texttt{pokemon} and recent releases like \texttt{helldivers}.
%
We find a thematic cluster of songs about technology. 
Especially for AI, there seems to be an equal number of songs in praise of AI and against it.
%
In the meme category we find clusters of songs featuring humorous, sometimes crass, lyrics or that feature words from popular online memes (e.g. \texttt{skibidi, sigma, Ohio}). 
%
In the ``other'' group, we find some clusters that are peculiar and some that seem to be artifacts of the clustering algorithm (highlighted by a question mark in the table).
In the former group, we have songs based on Edgar Allan Poe's poem \textit{The Raven}, which appear to be related to a community-organized challenge on the Suno Discord Server\footnote{\url{https://discord.com/channels/1069381916492562582/1231954456007151736/1253680890593280062} Last accessed Mar. 12 2025}. 
An example of the latter is in the cluster labeled \texttt{gpt glitch}, featuring lyrics that look like the output of an LLM, containing text such as \texttt{``I'm sorry but I can't create that for you.''}.
%

Manually inspecting the outliers shows that most of them are spread among existing clusters and could reasonably be merged.
Most notably, most of the points in the large outliers patch to the bottom left of the ``fantasy'' cluster could be part of it. Likewise, there is a dense group of outliers below the dance-related clusters that share the same semantic content.
There are however a few small clumps of points labeled outliers that are distinct groups.
For example, next to the other mixed languages clusters, there seems to be an undetected Arabic cluster.
Additionally, some of the outliers in this area feature either ASCII art or emojis in the lyrics.
In the upper part of the plot, there seems to be a certain number of sea shanties, mainly using ``The Wellerman'' lyrics.
On the right side there is a very dense group of songs with the same lyrics, featuring the line ``love is electric'', mostly created by the same user (found with manual inspection).

\subsection{Tags}\label{sec:tags}
\begin{figure}[h]
    \centering
    \includegraphics[width=1.0\linewidth]{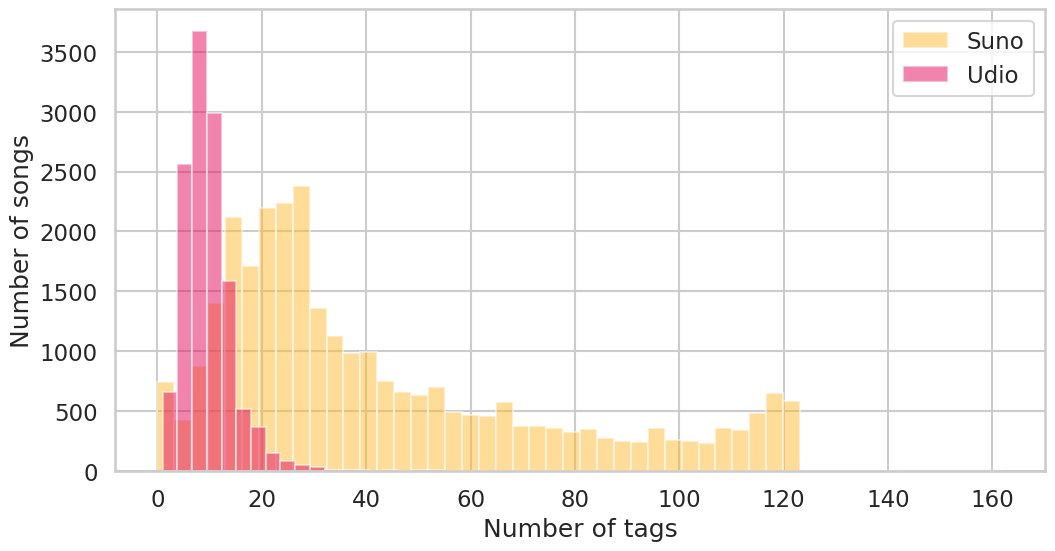}
    \caption{Length of the tags field in characters for Suno (yellow) and Udio (red). Suno tags are usually longer and more discursive compared to Udio.}
    \label{fig:tags-len}
\end{figure}

Tags contain genre and style conditioning for the generation model and the two platforms handle them slightly differently.
In Suno, they tend to be longer and more free in form, while in Udio, they tend to be shorter and formed by a list of comma-separated descriptors.
This is reflected in the length of tags histogram in Fig. \ref{fig:tags-len}.
The difference might be due to the way tags are collected from the interface of each service and by the practices that arise in each user community. 
In Udio the users seem to gravitate towards the use of comma-separated tags, with the interaction encouraging the use of multiple descriptors that get suggested adaptively.
In Suno, users seem to go for more free form descriptions, with the textbox being labeled `Style of Music', rather than tags. The interface however also offers clickable descriptors that are added as comma-separated elements.


\begin{figure*}[htb]
    \centering
    \includegraphics[width=0.43\linewidth]{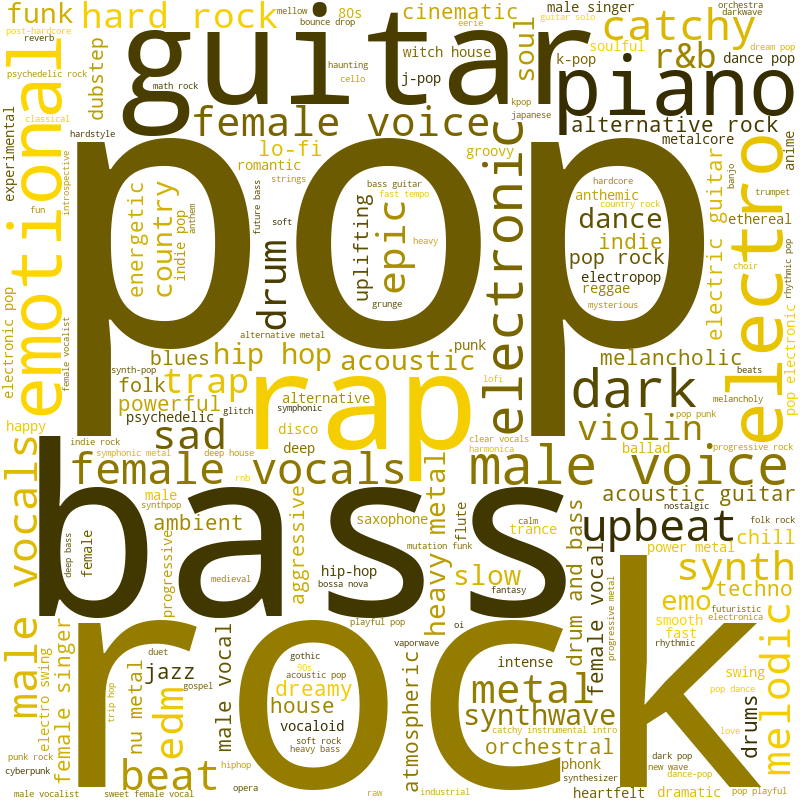}
    \hspace{1.5cm}
    \includegraphics[width=0.43\linewidth]{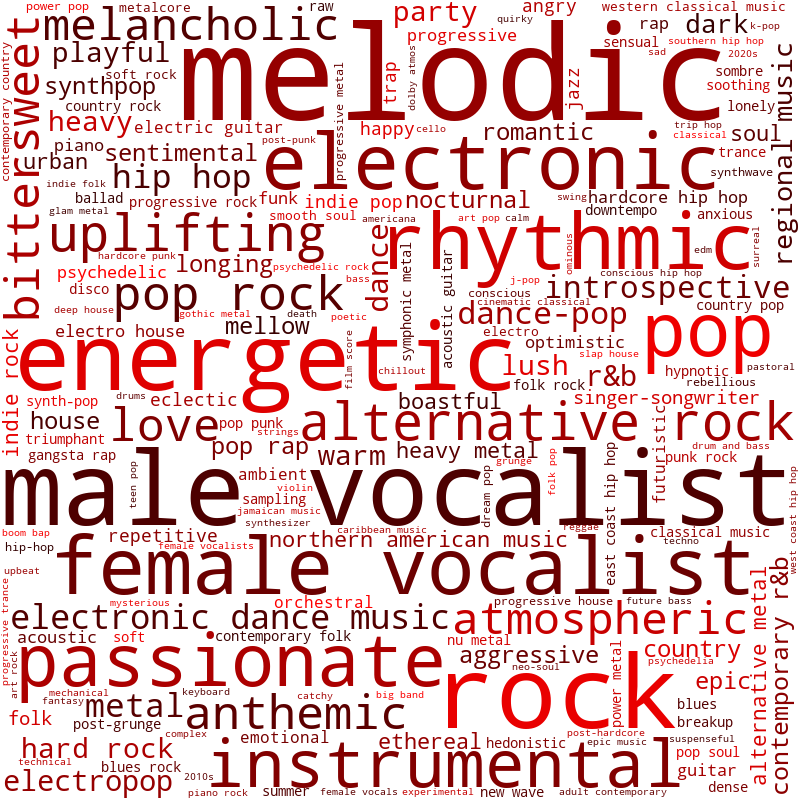}
    \caption{Word cloud for Suno (left) and Udio (right). Font size is scaled according to prevalence.}
    \label{fig:wordcloud}
\end{figure*}

For this analysis, we look at metadata labeled as \texttt{tags} in both platforms without attempting to reverse engineer the way they are collected or created.
In our dataset, we proceed to split the tag string by comma and strip extra symbols like parentheses.
After that, we count the prevalence of each individual tag and identify the most common.
Figure \ref{fig:wordcloud} shows a word cloud with the $200$ most common tags for each platform.
While we count $35,746$ unique tags, $80.7\%$ of them only appear once as they are either very uncommon words or very long descriptive strings.
From this point onward, we limit our analysis to tags that appear at least ten times in the dataset, resulting in $1,193$ tags.
We avoid reducing the strings to word stems to preserve individual tags in the comma-separated lists and avoid losing information from named entities like genre names.

We clustered the tags directly, but the resulting groups were not very informative, as semantic relationships seem to overpower functional ones.
For example, \texttt{guitar}, \texttt{rock} and \texttt{aggressive} would gravitate towards each other rather than be closer to instruments, genres and adjectives respectively.
To address this, we create a high-level taxonomy to guide the dimensionality reduction and influence the subsequent clustering. 
The classes are the following: Genre/Style; Instrument; Qualifiers/Mood; Structure; Voice; BPM; Key; Year/Decade; and Tempo.

We match genre names to those given by \textit{Every Noise at Once}.\footnote{\url{https://everynoise.com}, last accessed Mar. 12 2025.}
We do the same thing for musical instruments using the list from the \textit{Institute of Musical Instrument Technology} website.\footnote{\url{https://www.imit.org.uk/pages/a-to-z-of-musical-instrument.html}, last accessed Mar. 12 2025.}
Using a part-of-speech tagger, we find single-word adjectives in the remaining tags.
The remainder of the set contains the other classes that we spot in the first clustering attempt plus some outliers we sort out manually.
We match voice tags to anything containing the words ``voice'', ``vocal'' and ``singer''.
Similarly, structural tags are all those that contain keywords for song sections like chorus or intro. Tempo tags could be a subset of this category but stand out by the fact that they all contain the word \texttt{tempo}.
``BPM'', years (e.g., ``60s'') and key signatures (e.g., ``D flat major'') 
are all matched using regular expressions.
We refine these classes manually 
after checking the clustering results, which only needs minor adjustments.

\begin{figure*}[htb]
    \centering
    \includegraphics[width=\linewidth]{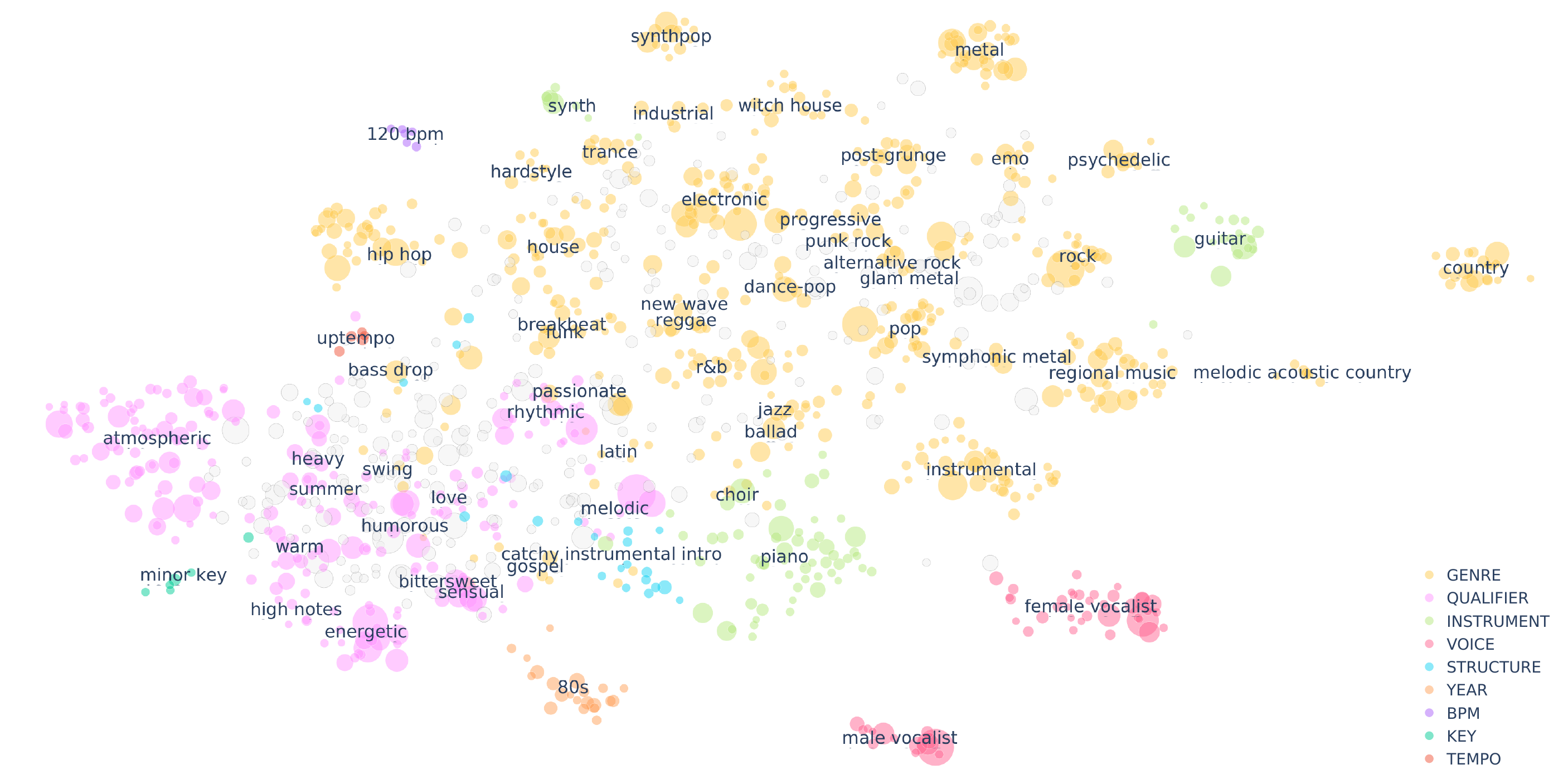}
    \caption{
        Clusters of the most common tags (combined ranking from both services).
        Colors correspond to macro categories defined manually.
        Text corresponds to the most prevalent tag in each cluster of the clusters we find with HDBSCAN. Grey indicates outliers.
    }
    \label{fig:tags_cluster}
\end{figure*}

We end up with $58$ clusters, mostly following the macro-categories we specify manually.
Figure \ref{fig:tags_cluster} shows the result of our clustering pipeline, where UMAP is conditioned on our high-level taxonomy.
The five-dimensional UMAP uses \texttt{n\_neighbors}$=15$ and \texttt{min\_dist}$=0.15$.
HDBSCAN uses minimum and maximum cluster sizes of $5$ and $50$ and $\epsilon=0.32$.
The two-dimensional UMAP uses \texttt{n\_neighbors}$=150$ and \texttt{min\_dist}$=0.45$.

Figure \ref{fig:tags_cluster} shows that genres, qualifiers, and instruments make up most of the dataset and are clearly separated.
Voice-related tags also form two compact groups of notable size for male and female voices, respectively.
We can see small, well-defined clusters for year, BPM and key specifiers.
Structural tags are the most spread out, as tags like \texttt{drop} and \texttt{beat} are pushed towards clusters featuring related genres.
Qualifiers are grouped into fewer and larger clusters, which appear to go from more abstract concepts (e.g., atmospheric) to more musical adjectives (e.g., \texttt{melodic}) that are closer to the genre clusters.
Genres are more easily separated and seem to cluster according to 
the similarity of the associated styles.
In some cases, like country or metal, the cluster is very compact and separated, as each entry is a combination of the genre name and an adjective (\texttt{black metal}, \texttt{thrash metal}, etc.). 
Instruments cluster together in three separate areas, with the guitar cluster lying next to rock and country, the piano cluster being closer to jazz and instrumental genres, and the synth cluster being closer to electronic music.
There are a number of elements that were not clustered and are represented in gray in Fig. \ref{fig:tags_cluster}.
Most of these outliers are quite prevalent tags that might however lie in between many different genres and concepts in the embedding space, e.g., \texttt{acoustic}, and are geometrically difficult to place.

\subsection{Real Artist Names in Udio}\label{sec:artists_names}
Udio replaces real artist names in the tags with a set of generic descriptors,
which might be a way for the company to avoid producing outputs that 
infringe intellectual property rights while retaining the ability for a user to
describe what music they want.
In the metadata, the \texttt{replaced\_tags} field is structured like a dictionary with labels for the nature of the replacement, making it easy for us to identify replaced artists.
In our collection of $20,519$ Udio songs, we find $1000$ such instances
corresponding to $703$ unique artists.
Table \ref{tab:udio_replaced} shows the $20$ most prevalent replaced artist names, 
along with the number of times they appear, and an example of the tags substituted for them.
Udio appears to match different spellings of the same name (e.g., Bjork or Bj\"ork) and the replaced tags are not always the same, suggesting a dynamic system behind the replacement and not just simply a lookup table.

\begin{table}[h]
\centering
\footnotesize
\begin{tabular}{m{1.5cm}cm{5.3cm}}
\toprule
Artist & \# & Replaced Tags \\ \midrule
XXXTentacion & 26 & emo rap, alternative r\&b, hip hop, contemporary r\&b, r\&b, pop rap, aggressive, self-hatred, boastful, depressive \\ \midrule
Drake & 19 & male vocalist, pop rap, contemporary r\&b, hip hop, r\&b, alternative r\&b, atmospheric, introspective, apathetic, mellow, bittersweet \\ \midrule
Taylor Swift & 18 & alt-pop, singer-songwriter, synthpop, nocturnal, romantic, love, atmospheric, lonely, sentimental, longing, concept album, lethargic, passionate, 2020s \\ \midrule
Foo Fighters & 18 & male vocalist, alternative rock, post-grunge, acoustic rock, energetic, melodic\\ \midrule
The Beatles & 17 & male vocalist, psychedelic pop, pop rock, psychedelia, sunshine pop, art pop, melodic, lush, love, fantasy, optimistic, dense, pastoral \\ \midrule
Depeche Mode & 17 & male vocalist, synthpop, downtempo, ambient pop, electronic, melancholic, melodic, calm, soothing, lush, mellow, nocturnal{]} \\ \midrule
Adele & 17 & female vocalist, pop soul, adult contemporary, pop, blue-eyed soul, passionate, sad, sentimental \\ \midrule
J. S. Bach & 16 & classical music, baroque music \\ \midrule
The Weeknd & 14 & male vocalist, alternative r\&b, electropop, r\&b, electronic, synthpop, nu-disco, party, hedonistic \\ \midrule
ABBA & 14 & female vocalist, europop, euro-disco, dance, pop, disco, optimistic, energetic, uplifting, melodic, rhythmic, party, lush\\ \bottomrule
\end{tabular}
\caption{The ten most replaced artists found in Udio's metadata under \textit{replaced\_tags}. The full list contains 703 artists.}
\label{tab:udio_replaced}
\end{table}

\subsection{Metatags in the Lyrics}\label{sec:ctrl}
Lyrics often contain elements that are not words to be sung by the model.
We will refer to these elements as \textit{metatags}, following a guide compiled by Suno user \textit{wetcircuit} found on \url{https://www.suno.wiki}.
Their purpose appears to be steering the music generation by providing additional information about characteristics like structure, instrumentation, delivery and dynamics.
Metatags mostly appear between square brackets, although they might also appear without special markings (e.g., {\tt Verse} followed by a new line) or using other symbols like parentheses.
We limit ourselves to looking for square brackets as it seems to be the dominant pattern.
Some user appear to put lyrical content like call and response phrases or second voices inside square brackets, these however drop to the bottom of our list since they only appear once.
Table \ref{tab:ctrl_counts} contains the 25 most common metatags and their prevalence in each platform.

\begin{table}[htb]
\centering
\footnotesize
\begin{tabular}{lrrr}
\hline
\toprule
\texttt{Sequence                } & Suno      & Udio      & Total \\ \midrule
\texttt{verse                   } & 73722     & 14548     & 88270 \\
\texttt{chorus                  } & 48670     & 16672     & 65342 \\
\texttt{bridge                  } & 17826     & 4665      & 22491 \\
\texttt{outro                   } & 6163      & 2386      & 8549  \\
\texttt{pre-chorus              } & 3725      & 2914      & 6639  \\
\texttt{end                     } & 3747      & 289       & 4036  \\
\texttt{intro                   } & 2436      & 1171      & 3607  \\
\texttt{instrumental            } & 2682      & 710       & 3392  \\
\texttt{drop                    } & 662       & 1403      & 2065  \\
\texttt{guitar solo             } & 1212      & 835       & 2047  \\
\texttt{hook                    } & 912       & 468       & 1380  \\
\texttt{break                   } & 943       & 212       & 1155  \\
\texttt{interlude               } & 538       & 398       & 936   \\
\texttt{fade out                } & 644       & 208       & 852   \\
\texttt{instrumental break      } & 504       & 280       & 784   \\
\texttt{solo                    } & 537       & 236       & 773   \\
\texttt{instrumental solo       } & 633       & 39        & 672   \\
\texttt{instrumental intro      } & 596       & 64        & 660   \\
\texttt{breakdown               } & 260       & 277       & 537   \\
\texttt{refrain                 } & 314       & 203       & 517   \\
\texttt{instrumental interlude  } & 452       & 55        & 507   \\
\texttt{yeah                    } & 3         & 466       & 469   \\
\texttt{pre-hook                } & 426       & 15        & 441   \\
\texttt{build                   } & 163       & 267       & 430   \\
\texttt{final chorus            } & 144       & 255       & 399   \\ \bottomrule
\end{tabular}
\caption{Top 25 most prevalent metatags. In cases where numbers appears, e.g., Chorus 2, those were stripped and merged into one category.}
\label{tab:ctrl_counts}
\end{table}

Most of the metatags we find are used to suggests structure,  with \texttt{verse} and \texttt{chorus} dominating the list.
Metatags for repeated parts often come with numbers, e.g., \texttt{chorus 2} but we opted to remove them and just consider the main metatag.
Words specifying the instrumentation for the section like \texttt{guitar solo} or \texttt{instrumental intro} were very common.

Some users go a step further by including multiple instructions in a single metatag.
Looking at the least frequent elements in the subset, we find some unique and very elaborate metatags.
A few examples of recurring elements are: chord progressions, precise duration, specific instruction on instrumentation, dynamics or tempo.
However, as noticed in \textit{suno.wiki}, long sequences of metatags seem to get ignored by the AI model and can be misinterpreted as lyrics and be sung instead.

The heterogeneity in the way these sequences are used suggests this is an evolving practice, with users experimenting and trying to build shared knowledge of how to best interact with the systems. 
Udio also differs from Suno in that they actively encourage users to experiment by providing a tool-tip with suggestions that the user can summon using the \texttt{/} key.
We can speculate that this behavior could be exploited by Udio to ultimately improve the platform's prompting capabilities by leveraging crowd-source labels.

Like we see with tags, sometimes metatags include the names of real artists.
This is especially important in Suno where providing such names in the tags is forbidden.
Table \ref{tab:ctrl_artists} contains a few randomly selected examples of such metatags, where the name of an artist is used in an attempt to evoke their voice or music style.

\begin{table}[h]
    \centering
    \footnotesize
    \begin{tabular}{lc}
    \toprule
        Artist & Metatag \\ 
        \midrule
        Bob Marley  & \texttt{produced by bob marley and lee perry}  \\
        Journey     & \texttt{journey separate ways synth arpeggio}  \\
        Kanye West  & \texttt{verse 2: kanye west}  \\
        Madonna     & \texttt{influence: madonna, michael jackson} \\
        \bottomrule
    \end{tabular}
    \caption{A few examples of metatags that mention real artists}
    \label{tab:ctrl_artists}
\end{table}

\section{Discussion}\label{sec:discussion}
We now critically discuss our work and point out limitations. 
The principal goal of this article is to present the first
systematic analysis of the users of AI music generation platforms
through their prompts and lyrics.
We focus our analysis on two specific popular platforms: Suno and Udio.
We aim to understand how these platforms are being used,
and what topics are inspiring their users.
To this end, we collect a large dataset of metadata from these platforms,
and build a processing pipeline and a set of interactive visualizations that allow us to identify and characterize prevalent semantic groups of prompts, tags and lyrics.

The resulting picture of AI music is of course limited by the six-month time frame of our data collection, namely May to October 2024.
Certain trends might not be captured, and the prevalence of certain themes or tags 
can be due to their temporal context.
For example, memes and geopolitical references shift continuously and sometimes very quickly.
Another aspect that can be important for some users but is not captured by our dataset 
is the presence of different versions for each service.
There can be interesting differences in the way different model versions respond to prompting, which in turn will be reflected in different prompting practices being used. 
Finally, our dataset consists only of songs published by users,
which inevitably introduces a bias on the quality of songs in the dataset, 
as well as the prevalence of certain themes. 
Unfortunately, we see no way around this limitation. 

Our focus on English is a clear limitation of this study,
and so future work should look for trends and peculiarities in other languages.
Cultural differences might emerge in the way users with different linguistic and national backgrounds engage with AI music generation.
An interesting aspect to consider is how the distribution of languages on the platforms does not seem to follow the distribution of the most commonly spoken languages in the world.
This might be due to asymmetries in the way the services were released and advertised across the world, as well as economic factors and internet access.

While the clustering methodology we employ is extremely effective at extracting information from such a large and heterogeneous collection,
it ends up labeling many points as outliers.
Manually inspecting these outliers, however, shows that there are undetected groups that could still be recovered by introducing a classifier trained on data that has been labeled and validated by humans.
Future work might consider a multi-step procedure that involves active learning to have total coverage.
Additionally, this would allow continuous addition of new songs to the dataset and automatic labeling of them.
While in our work we named clusters using a coarse procedure that required some manual refinement, 
a more sophisticated pipeline could be employed, possibly including large language models for naming the clusters. 

While performing this research, we often found ourselves at methodological crossroads.
On the one hand, we want to capture the most representative user behaviors and song themes in our dataset, giving special attention to establishing well-defined clusters and analyzing the major ones.
On the other hand, outliers stick out from the bulk of the data and contain interesting behaviors that can reveal something about the more nuanced inner workings of the platforms and the environment around them.
We try to strike a balance in this article, but future work might focus on specific thematic or linguistic subgroups identifiable with our methodology.

Ethnographic studies can be carried out targeting specific user groups and their unique practices.
AI music communities can also be studied by integrating relevant data from Discord or Reddit into the analysis.
General discussions in those forums might provide insight on how AI music is received and what strategies and practices are applied in the process and are considered the best.
For example, studying challenges organized by community members can provide multiple songs with a common thematic baseline to be analyzed, helping highlight differences.


\section{Conclusion}\label{sec:conclusion}
This article is the first to study textual metadata from a large dataset of AI music collected from two platforms offering text-to-music models: Suno and Udio.
The dataset we collect is the first one of its kind, 
and allows us to look at the growing phenomenon of AI music.
We draw from existing literature on AI-image generation, and we adapt an established methodology to study prompt-based systems.
We focus on prompts, tags and lyrics as they are the main interface to these music generation systems and provide insight into the users of the platforms and how they use them.
Our analyses identify clusters of themes in the data and highlighting behaviorism and trends in the prompts and tags of the users.
Due to uncertainty regarding the sharing of our dataset
with regards to intellectual property rights,
we only make available a list of URLs with which one can retrieve the data.
Our novel methodology and data can be used by other researchers in the field that are interested in furthering this exploration.
We believe that a number of relevant downstream tasks can be based on our data and methods.
Ethnographic studies might target subgroups in the user base and studying specific aspects that we only touched upon like non-English prompts and lyrics, while more technical research can benchmark new MIR pipelines on the large dataset.
In the future, we plan on extending on this work by leveraging both audio and textual features and their interplay to support further analyses.

\section{Reproducibility}\label{sec:reproduce}
We provide the code we use to gather the data and to produce our visualization and analysis at the following repository \url{https://github.com/mister-magpie/aims_prompts}.
For legal concerns we do not publish our dataset with the song metadata.
However, we include a list of URLs to the songs we collected from both services.
The availability of songs is subject to changes as users can decide to delete their songs or accounts.
At the time of submission we observed that $4.59\%$ and $1.02$ of URLs from Udio and Suno respectively are not reachable.


\section{Ethics and consent}\label{sec:ethics}
As AI generated media and services used to produce it become more widespread,
it is important to study it and its impact \cite{sturm2024ai}.
Since this technology exists in the public sphere it should be studied
in this real context with real users.
Our research questions revolve around users and their use, 
and so we focus on observing AI music as users have created and published it
over a six month time period.
Our collection of data from the Suno and Udio website are 
not permitted by their terms of service;
but that does not mean this research is illegal or unethical.
As we are a publicly funded research group focused on understanding
the nature of AI music and its broader implications, we are permitted
to collect such data and study it under the Copyright Directive (EU) 2019/790 \cite{EUDirective2019}.

\section*{Acknowledgements}
This work was supported by the grant ERC-2019-COG No. 864189 \href{https://musaiclab.wordpress.com/}{MUSAiC: Music at the Frontiers of Artificial Creativity and Criticism}, 
and by the Wallenberg AI, Autonomous Systems and Software Program – Humanities and Society (WASP-HS) funded by the Marianne and Marcus Wallenberg Foundation (Grant 2020.0102)


\printbibliography

\end{document}